# A Bayesian baseline for belief in uncommon events

V. Palonen

Department of Physics, University of Helsinki, P.O. Box 43, 00014 University of Helsinki, Finland

vesa.palonen@helsinki.fi

## ABSTRACT

The plausibility of uncommon events and miracles based on testimony of such an event has been much discussed. When analyzing the probabilities involved, it has mostly been assumed that the common events can be taken as data in the calculations. However, we usually have only testimonies for the common events. While this difference does not have a significant effect on the inductive part of the inference, it has a large influence on how one should view the reliability of testimonies. In this work, a full Bayesian solution is given for the more realistic case, where one has a large number of testimonies for a common event and one testimony for an uncommon event. It is seen that, in order for there to be a large amount of testimonies for a common event, the testimonies will probably be quite reliable. For this reason, because the testimonies are quite reliable based on the testimonies for the common events, the probability for the uncommon event, given a testimony for it, is also higher. Hence, one should be more open-minded when considering the plausibility of uncommon events.

# INTRODUCTION

Is it reasonable to believe in a testimony of an uncommon event in the face of uniform contrary evidence from prior events? This question has been much discussed historically, with notable contributions from David Hume (Hume 1748) , John Earman (Earman 2000), Millican (Millican 2013), and many others.

David Hume's argument was not clearly formulated, but Hume basically argued that the evidence for common events is so strong that the testimony for uncommon events (miracles) is usually not strong enough evidence for the uncommon event to be believable. "Extraordinary claims require extraordinary evidence" is the oft-used and dangerously poorly-defined phrase often connected to Hume's position.

Earman does a systematic job of both trying to find a precise form for Hume's argument and then showing the problems thereof. Earman makes two important points:

1. With a Bayesian calculation of inductive inference, the probability of an uncommon event does indeed go down with the amount of common events (as $1/(n+2)$), but never to zero. Hence, based on induction, one can hence never be certain that the uncommon won't happen.
2. Earman discusses the role and reliability of testimonies for uncommon events. Earman shows that the testimony can often provide enough credibility for the uncommon event. Notably, in considering the evidential force of a testimony, one needs to consider, not just how often witnesses are wrong in general, but what is the probability that the witness would make just such a particular claim and be in error with that claim.  For example, when a witness is testifying that John Doe won the lottery, it is not enough to suggest that a testimony is in general wrong with e.g. 10% probability, but one needs to take into account the probability that the claim was made about John Doe in particular (why just him?) and also what is the probability that the claim indeed would be erroneous.

Presently the calculations published on the topic assume a large amount of common events. In reality we usually only have a large amount of testimonies for the common events. That is, we do not have a uniform *evidence* against the uncommon events. What we may have is uniform *testimony* against the uncommon events or miracles. In this sense it can be said that up to now the problem of uncommon events and their believability based on testimony has not been fully analyzed even on the basic level.

This paper offers a full Bayesian solution for the more realistic case, namely, for the question: How believable is an uncommon event when we have a uniform mass of *testimony* to the contrary? The calculation can be seen as a baseline for further discussions on the topic, with nuances to be added as later as different additions and changes to the model are explored. Further consideration will involve considerations for several testimonies of the same event, the independence of the witnesses, the effects of prior beliefs against the uncommon event, and whether or not those testifying to a rare event are less trustworthy than those testifying to a common event.

a further complication in the field has been that the usage of probabilities in the discussion has been partial, with several authors dissecting the full formulas for partial arguments based on the full formulation, see e.g. (Ahmed 2015), with the full solution nowhere to be seen. The aim here will be to show the full solution for the simple default situation with few assumptions. From there, different assumptions can be added whenever the assumptions are well grounded.

The calculation will be made for a general case for which we have testimonies of common events and one testimony of an uncommon event. The results will then apply to miracles, testimonies of rare natural events like winning a lottery, and rare-event measurements in physics (e.g. proton decay).

# NOMENCLATURE

Below is a table of notations used in the paper. For simplicity, the logical and symbol ∩ is usually dropped in the probability notation.

| | |
|---|---|
| $\neg A$ | Not A |
| $A \cap B$ | A and B |
| $A \cup B$ | A or B |
| $p(A \mid B)$ | Conditional probability of A being true given that B is true |
| $p(A\,B)$ | Probability of A and B. |
| $W$ | An uncommon event (white ball drawn from an urn) |
| $B$ | A common event (black ball drawn for an urn) |
| $B_i$ | The result of the *i*'th event is common (*i*'th ball was black) |
| t(…) | Testimony of an event |
| $n$ | Number of testimonies of a common event |
| $t(B)^n$ | *n* testimonies of a common event |
| $C^n$ | A vector of *n* real events (W or B) behind the *n* testimonies |
| $v$ | The (unknown) probability for the uncommon event to happen |
| $d$ | The (unknown) probability for a testimony to be wrong, $d = p(W|t(B)) = p(B|t(W))$. |

# THE BASELINE MODEL

Using the above notation, the simple general case is this: There are *n* testimonies t of a common event B, $t(B)^n$ and one testimony for an uncommon event W, $t(W)$. What is the probability of W being in fact true given the testimonies, $p(W \mid t(W)\, t(B)^n)$?

We will assume as little as we can about the reliability of the witnesses (d) and about the real probability of the uncommon event happening (v). In effect, we will assign only reasonable prior probabilities for these probabilities and in the end let the data decide the most probable values for these probabilities. (These kinds of priors are often called hyperpriors in the data-analysis literature.) For simplicity, we will use uniform priors

$$v \sim U(0, 1),$$

$$d \sim U(0, 0.2),$$

where the notation $x \sim U(a, b)$ means that the probability density for *x* is constant between *a* and *b* and zero elsewhere. With the latter prior we have assumed that in general the testimonies are over 80% reliable, an assumption which will be seen to matter less and less as *n* increases.

We will be using general Bayesian methodology, which is basically finding out the joint probability distribution for all the parameters relevant to the case and calculating the wanted probability distribution from the joint distribution by using marginalization and the Bayes rule. (This approach is generally applicable and much used in the machine-learning community because from the joint distribution one can systematically calculate whatever probability one happens to need.)

In this case, the joint distribution factors as (see Appendix A for details)

$$p(W\ C^n\ t(W)\ t(B)^n\ v\ d) = p(v)\ p(d)\ p(W|v)\ p(C^n|v)\ p(t(W)\ |\ W\ d)\ p(t(B)^n|C^n d)$$

And the wanted probability is

$$p(W\ |\ t(W)\ t(B)^n) = \frac{p(W\ t(W)\ t(B)^n)}{p(t(W)\ t(B)^n)} = \frac{p(W\ t(W)\ t(B)^n)}{p(W\ t(W)\ t(B)^n) + p(B\ t(W)\ t(B)^n)} = \frac{J_W}{J_W + J_B}$$

where we have terms of the form (by marginalization)

$$J_W = p(W\ t(W)\ t(B)^n) = \sum_{\forall C^n} \int_0^1 dv \int_0^{0.2} dd\ p(W\ C^n\ t(W)\ t(B)^n\ v\ d)$$

where the sum is over all possible combinations of the elements of $C^n$, that is, we marginalize over all the possibilities in $(W_1 \cup B_1) \cap (W_2 \cup B_2) \cap \ldots \cap (W_n \cup B_n)$. After calculations, the terms amount to (see the Appendix A for details)

$$J_W = \int_0^1 dv \int_0^{0.2} dd\ v(1-d)(2dv - d - v + 1)^n$$

and

$$J_B = \int_0^1 dv \int_0^{0.2} dd\ (1-v)\ d\ (2dv - d - v + 1)^n$$

With these terms in hand, we are now in the position to show some results.

## Results of the baseline model

To reiterate, in previous works (see e.g. (Earman 2000)), it has been shown that when *n* common events are taken as data, simple Bayesian inference with reasonable priors assigns a 1/(*n*+2) probability for the uncommon event happening. This simple case of inductive inference does not take into account the testimonies for the events (common or uncommon), as is done in the current model.

Figure 1 shows the probability for the uncommon event, with one testimony for the uncommon event, as a function of *n*, the number of testimonies for the common event, $p(W\ |\ t(W)\ t(B)^n)$. Perhaps surprisingly, as the number of testimonies for the common event (*n*) grows large, the probability for the uncommon event given the testimonies approaches the value 0.5 asymptotically. There is a very large difference to the results of the simple inductive inference mentioned above, where the probability approaches zero asymptotically.

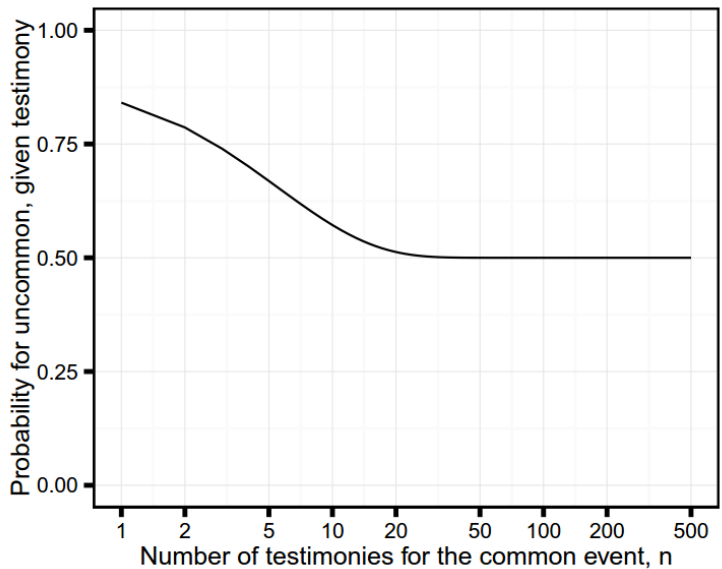

**Figure 1. Probability for the uncommon event in the face of n testimonies for a common event. Note the logarithmic horizontal axis.**

What is the reason for the difference of the results for the present more realistic model? Why does even one testimony for an uncommon event overcome the inductive part of the inference from the large amount of common events?

The basic reason is that, for there to be a large consistent amount of testimonies for the common events, the testimonies themselves have to be reliable. That is, if the testimonies were unreliable, it would be unlikely to have a uniform set of testimonies for the common case. Rather, there would likely be some testimonies for the uncommon event. On the other hand, if there are some past testimonies for the uncommon event, the inductive part of the inference will not be so strong against the uncommon events. Figure 2 shows the mean values of the probability of the uncommon event happening ($v$) and of the probability of a testimony being false ($d$). It is seen that as the number of testimonies ($n$) for the common event increases, the probability of the uncommon event decreases as expected, but at the same time the probability of a false testimony also decreases, and roughly at the same rate. Hence, even one testimony for an uncommon event is able to balance out the inductive part of the inference and make the uncommon event believable.

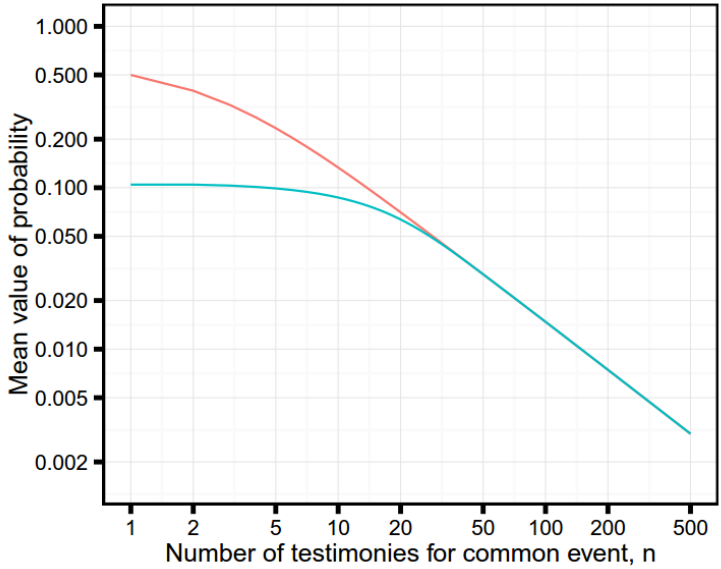

**Figure 2. Mean values for the probabilities for the uncommon event (red) and false testimony (blue). Note the logarithmic axes.**

## APPENDED CASE WITH KNOWN ERRONEOUS TESTIMONIES

Let us now append the previous case by including an *l* amount of false testimonies for the uncommon event. Our additional data is then $B^l \cap t(W)^l$. The probability we will be interested in is $p(W \mid t(W)\ t(B)^n\ B^l\ t(W)^l)$.

The joint distribution will now factor as (see Appendix B for more details)

$$p(W\ C^n\ t(W)\ t(B)^n\ B^l\ t(W)^l\ v\ d)$$
$$= p(v)\ p(d)\ p(W|v)\ p(C^n|v)p(B^l|v)\ p(t(W) \mid W\ d)\ p(t(W)^l|B^l\ d)\ p(t(B)^n|C^n\ d)$$

The calculations will proceed as before, with some additional term. The wanted probability is again of the form

$$p(W \mid t(W)\ t(B)^n\ B^l\ t(W)^l) = \frac{J_W'}{J_W' + J_B'}$$

where

$$J_W' = \int_0^1 dv \int_0^{0.2} dd\ v(1-v)^l(1-d)d^l(2dv - d - v + 1)^n$$

$$J_B' = \int_0^1 dv \int_0^{0.2} dd\ (1-v)^{l+1}\ d^{l+1}\ (2dv - d - v + 1)^n$$

## Results for the case with erroneous testimonies

Figure 3 shows the probability for the uncommon event given the testimonies for the appended case. Shown are cases with the number of known false testimonies $l$ = 0, 1, 3, 10, and 50.

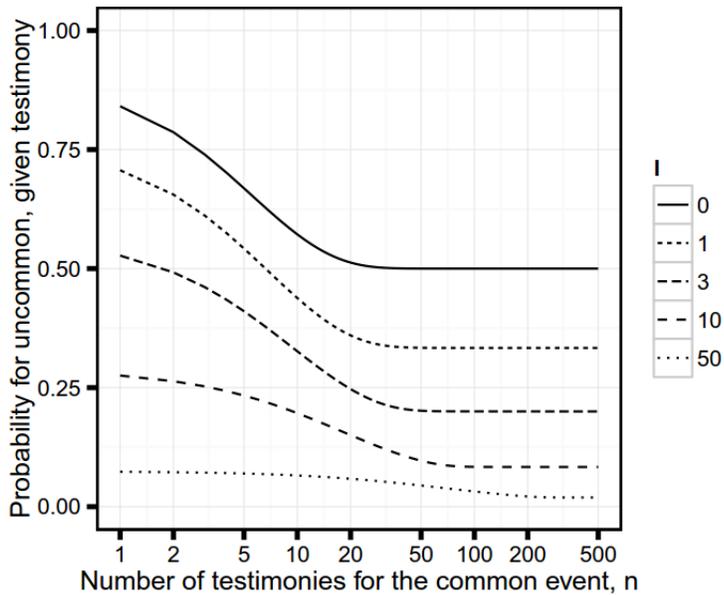

**Figure 3. Probability for the uncommon event in the face of n testimonies for a common event given different amounts of known false testimonies. Note the logarithmic horizontal axis.**

Figure 4 shows the mean values of the probabilities of the uncommon event happening ($v$) and for a testimony being false ($d$) for cases with different number of known false testimonies for the uncommon event.

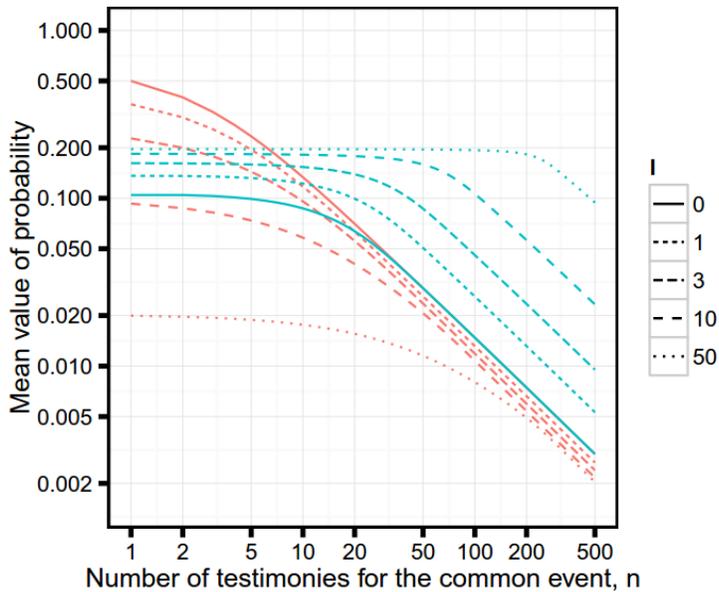

**Figure 4.** Mean values for the probabilities for the uncommon event (red) and false testimony (blue), given a number l of known false testimonies. Note the logarithmic axes.

One can see from the results that a small amount of known false testimonies for uncommon events does not significantly alter the believability of an uncommon event for which one testimony is not known to be false. For example, with three known false testimonies for an uncommon event and a large number of testimonies for common events, the probability for an uncommon event given one testimony for it is still roughly 0.2.

# CONCLUSIONS

The main result of the paper is that, when we have a large amount of testimonies for a common event and even only one testimony for an uncommon event, the probability we should assign for the uncommon event surprisingly large, namely 0.5. This is assuming that, without information to the contrary, we are treating all the testimonies the same way, and we are not assuming additional structure (model comparison) for reality behind the events.

This result has relevance for the study of miracles and also for science. In science, we should be more open to testimonies for "weird" empirical results which may not fit the current theoretical understanding. For example, Dr. Daniel Shechtman's discovery of quasicrystals (Shechtman et al. 1984) should have been met with more of an open mind by the community.

In the case with some known-false testimonies for the uncommon event, the probability for the uncommon event is lower but not significantly so. Hence, the additional Humean argument against uncommon events based on some false testimonies of uncommon events does not seem to have much force.

It is noted that in the present model very few assumptions were made and e.g. the probabilities for an uncommon event and the testimonies were left open and decided based on the available data. Yet, and importantly, it was assumed that the probability of a false testimony is symmetric, that is, that it is as likely for a person to make a mistake in the testimony for an uncommon as in the testimony for a common event. Hence, the number of testimonies for a common event had a bearing on the reliability on testimonies in general and hence also for the testimony for the uncommon event. It might be tempting to disconnect the two probabilities or to assume that a testimony for an uncommon event is more likely to be false than a testimony for a common event. While the former is possible, it would be hard to maintain that there is no connection between the reliability for the testimonies of uncommon and common events, the disconnection possibly leading to absurd results for low values of *n*. The latter option of assuming that the testimonies for uncommon events are less reliable seems biased. Because such an assumption would equate bringing more information to bear on the case, there should be a clear and agreed-on grounding for making this assumption. The author suspects that such an assumption is not sustainable, but leaves that for further, more nuanced, discussions.

For the model with known false testimonies for the uncommon event, the false testimonies might be viewed as a reason to relax the symmetry of the reliability of the testimonies of common and uncommon events. This exercise and grounding thereof is also left for further study on the matter.

# REFERENCES

Ahmed, Arif. 2015. "Hume and the Independent Witnesses." *Mind* 124 (496) (October 4): 1013–1044. doi:10.1093/mind/fzv076.

Earman, john. 2000. *Hume's Abject Failure : The Argument Against Miracles*. Oxford University Press, USA.

Hume, David. 1748. *An Enquiry Concerning Human Understanding*. London: A. Millar. https://ebooks.adelaide.edu.au/h/hume/david/h92e/.

Millican, Peter. 2013. "Earman on Hume on Miracles." In *Debates in Modern Philosophy: Essential Readings and Contemporary Responses*, edited by Stewart Duncan and Antonia LoLordo. Routledge.

Neapolitan, Richard E. 2004. *Learning Bayesian Networks*. Pearson Prentice Hall.

Pearl, Judea. 1997. "Bayesian Networks." *UCLA Computer Science Department, Technical Report* R246: 1–5.

Shechtman, D., I. Blech, D. Gratias, and J. W. Cahn. 1984. "Metallic Phase with Long-Range Orientational Order and No Translational Symmetry." *Physical Review Letters* 53 (20) (November 12): 1951–1953. doi:10.1103/PhysRevLett.53.1951.

# APPENDIX A DETAILED CALCULATIONS FOR THE BASELINE MODEL

## Joint factorization

Figure A1 gives the dependencies between the parameters of the model as a directed acyclic graph (Pearl 1997; Neapolitan 2004).

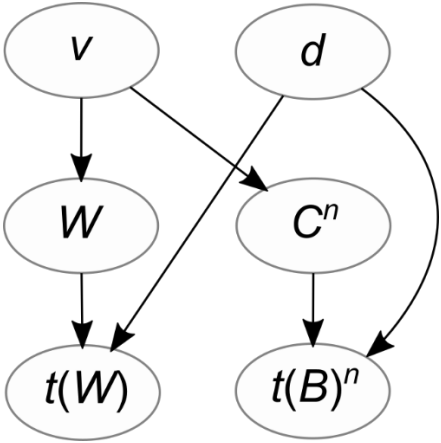

**Figure A1. Directed acyclic graph of the case.**

The arrows in the graph represent direct probabilistic dependencies between the parameters of the model. The natural factorization of the joint distribution can be read from the DAG (Neapolitan 2004) to be

$$p(W\ C^n\ t(W)\ t(B)^n\ v\ d) = p(v)\ p(d)\ p(W|v)\ p(C^n|v)\ p(t(W)\ |\ W\ d)\ p(t(B)^n|C^n d).$$

## Summation over possibilities of $C^n$

Recall that in the simple model we have two terms of the form

$$J_W = p(W\ t(W)\ t(B)^n) = \sum_{\forall C^n} \int_0^1 dv \int_0^{0.2} dd\ p(W\ C^n\ t(W)\ t(B)^n\ v\ d)$$

In this section we will calculate this term, notably the sum over all possibilities of $C^n$. Now

$$J_W = \sum_{\forall C^n} \int_0^1 dv \int_0^{0.2} dd\ p(v)\ p(d)\ p(W|v)\ p(C^n|v)\ p(t(W)\ |\ W\ d)\ p(t(B)^n|C^n d)$$

$$= \int_0^1 dv\ p(v)\ p(W|v) \int_0^{0.2} dd\ p(d)\ p(t(W)\ |\ W\ d) \sum_{\forall C^n} p(C^n|v)\ p(t(B)^n|C^n d)$$

$$= c \int_0^1 dv\ p(W|v) \int_0^{0.2} dd\ p(t(W)\ |\ W\ d)\ S_n\ ,$$

where the constant $c$ is a product of the constant priors of $v$ and $d$, and

$$p(W|v) = v$$

$$p(t(W) \mid W\ d) = 1 - d$$

$$S_n = \sum_{\forall C^n} p(C^n|v)\, p(t(B)^n|C^n d) = (1 - v - d + 2vd)^n$$

The following is an inductive proof for the last identity

For $S_2$, the sum is over the possibilities $(W_1 \cup B_1) \cap (W_2 \cup B_2)$

$$\begin{aligned}S_2 &= p(W_1\,W_2 \mid v)\, p(t(B_1)t(B_2) \mid W_1\,W_2\,d) + p(W_1\,B_2 \mid v)\, p(t(B_1)t(B_2) \mid W_1\,B_2\,d) \\ &\quad + p(B_1\,W_2 \mid v)\, p(t(B_1)t(B_2) \mid B_1\,W_2\,d) + p(B_1\,B_2 \mid v)\, p(t(B_1)t(B_2) \mid B_1\,B_2\,d) \\ &= v^2 d^2 + 2v(1-v)d(1-d) + (1-v)^2(1-d)^2 = (1 - v - d + 2vd)^2.\end{aligned}$$

Next, with a lower case $c_i$ we will denote the $i$'th element of $C^n$ and similarly for $t(B)^n$. For $S_{n+1}$, we have

$$\begin{aligned}S_{n+1} &= \sum_{\forall C^{n+1}} p(C^{n+1}|v)\, p(t(B)^{n+1}|C^{n+1}d) = \sum_{\forall c_{n+1}} \sum_{\forall C^n} p(C^n c_{n+1}|v)\, p(t(B)^n t(B)_{n+1}|C^n c_{n+1} d) \\ &= \sum_{\forall c_{n+1}} \sum_{\forall C^n} p(C^n|v)\, p(c_{n+1}|v)\, p(t(B)^n|C^n d)\, p(t(B)_{n+1}|c_{n+1} d) \\ &= S_n \sum_{\forall c_{n+1}} p(c_{n+1}|v) p(t(B)_{n+1}|c_{n+1} d) = (1 - v - d + 2vd)^{n+1}\ \blacksquare\end{aligned}$$

# APPENDIX B DETAILED CALCULATIONS FOR THE CASE WITH ERRONEOUS TESTIMONIES

Figure B1 gives the dependencies between the parameters of the model.

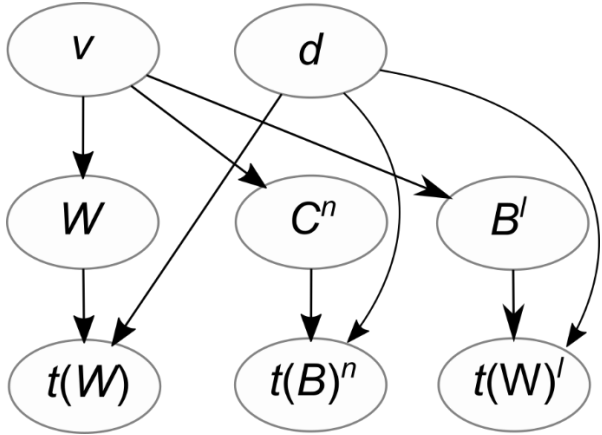

**Figure B1. A directed acyclic graph of the model with erraneous testimonies.**

Again, the joint distribution can be read from the graph to be

$$p(W\ C^n\ t(W)\ t(B)^n\ B^l\ t(W)^l\ v\ d)$$
$$= p(v)\ p(d)\ p(W|v)\ p(C^n|v)p(B^l|v)\ p(t(W)\ |\ W\ d)\ p(t(W)^l|B^l\ d)\ p(\ t(B)^n|C^n\ d)$$

And the wanted probability is

$$p(W\ |\ t(W)\ t(B)^n\ B^l\ t(W)^l) = \frac{p(W\ t(W)\ t(B)^n\ B^l\ t(W)^l)}{p(t(W)\ t(B)^n\ B^l\ t(W)^l)} = \frac{J_W{'}}{J_W{'} + J_B{'}}$$

Where

$$J_W' = \sum_{\forall C^n} \int_0^1 dv \int_0^{0.2} dd\ p(W\ C^n\ t(W)\ t(B)^n\ B^l\ t(W)^l\ v\ d)$$

$$= \int_0^1 dv\ p(v)\ p(W|v)p(B^l|v) \int_0^{0.2} dd\ p(d)\ p(t(W)\ |\ W\ d)\ p(t(W)^l|B^l\ d) \sum_{\forall C^n} p(C^n|v)\ p(\ t(B)^n|C^n\ d)$$

$$= c \int_0^1 dv\ v(1-v)^l \int_0^{0.2} dd\ (1-d)d^l\ S_n.$$

And similarly

$$J'_B = c \int_0^1 dv\ (1-v)^{l+1} \int_0^{0.2} dd\ d^{l+1}\ S_n.$$